\documentclass[12pt]{article}
\usepackage{amssymb,latexsym}
\usepackage{amsmath}

\columnwidth\textwidth

\begin{document}

\newcommand{\be}{\begin{equation}}
\newcommand{\ee}{\end{equation}}
\newcommand{\beann}{\begin{eqnarray*}}
\newcommand{\eeann}{\end{eqnarray*}}
\newcommand{\bea}{\begin{eqnarray}}
\newcommand{\eea}{\end{eqnarray}}
\newcommand{\nn}{\nonumber}
\newtheorem{df}{Definition}
\newtheorem{thm}{Theorem}
\newtheorem{lem}{Lemma}

\begin{titlepage}

\noindent
\vspace*{1cm}
\begin{center}
{\LARGE Liberties in Nature. On Photons, Bugs and Chess Players}

\vspace{2cm}

P. H\'{a}j\'{\i}\v{c}ek \\
Institute for Theoretical Physics \\
University of Bern \\
Sidlerstrasse 5, CH-3012 Bern, Switzerland \\
hajicek@itp.unibe.ch
\\ \vspace{1.5cm}

November 2006  \vspace*{1cm}

\nopagebreak[4]

\begin{abstract}
  Free will is an old philosophical enigma that has been recently revived by
  neuropsychology. We restrict ourselves to the problem that determinism seems
  to allow only an illusion of freedom but random decissions do not contain
  any freedom either. We show that this is a problem of natural sciences, not
  philosophy. Physics motivates replacing determinism by the principle of weak
  causality and introducing the concept of liberty. Its empirical basis
  remains untouched, but the theoretical interpretations of the state space in
  Newton theory and of the space-time in general relativity are changed. The
  emerging understanding of time agrees with the idea suggested once by
  Popper. In biology, the most important liberties are those of mutation, of
  motion and of the portable neural representation. We distinguish freedom and
  liberty. Each freedom is associated with some liberty and is defined as the
  ability to perform three processes called realization, selection and use of
  memory. This makes freedom accessible to experimental study. The freedom of
  will is explained by giving account of the underlying specific liberty and
  processes. While the realization has an ample space for randomness, the
  selection is mostly causal. Thus, determinism can be rejected without
  forcing decissions to be random. The experiments of Libet and the role of
  consciousness are discussed.
\end{abstract}

\end{center}

\end{titlepage}

\section{Introduction}
The interest in the free will has been enhanced today by the progress in
neuropsychology. \cite{Rama}. The old philosophical enigma is concisely
described in \cite{Pink}. One particular problem seems to be the following. If
determinism is assumed, then the free will appears to be an illusion:
everything is fixed and there seems to be no freedom.\footnote{There is a
  philosophical school called compatibilism trying to define freedom in a way
  compatible with determinism \cite{Dennett}. It is not satisfactory because
  the assumption of determinism is not plausible.}
On the other hand, if determinism is rejected, then it seems that decissions
of the will must be random. Thus, there is no control and this does not look
very free either. There are other problems associated with the free will, such
as morality etc. We shall ignore these here.

In the present paper, we are going to explain how a solution of the above
problem emerges from natural sciences. Physics teaches us how determinism can
be abolished and biology shows the way of how decissions of the will then
still need not become random. The paper is necessarily a sketch of a research
project rather than a complete scientific analysis considering all details and
aspects. But we shall try hard to formulate the main ideas as clearly and
distinctly as possible and to make them accessible to readers with various
backgrounds.

The plan of the paper is as follows. Sec.\ 2 is dedicated to physics and asks
the question which physical phenomena are strictly tied by unique rules and
which liberty remains after the physical laws are accepted. We find that the
liberty is large even in Newton mechanics: the choice of system and the choice
of its initial data are free. We give this liberty a slightly different
interpretation than is usually adopted. Then, a short account of quantum
mechanics will show that it allows even more liberty and that the additional
liberty does not concern exclusively the micro-world. This motivates the
formulation of the weak causality principle and the corresponding notion of
time on which the conceptual framework of this paper is based (c.f.\
\cite{Popper1}). This notion of time is not compatible with the usual
understanding of general relativity, but a subtle change in the interpretation
of spacetime can make the framework logically coherent. This reinterpretation
does not influence any observable property.

The examples met in physics motivate the introduction of the central notion of
the paper, the liberty, in Sec.\ 3. A liberty is constituted by alternative
possibilities under certain conditions and is experimentally testable.

The final section lists the most important liberties that concern living
organisms. It is the liberty of mutation, of motion and of portable neural
representative. While the first two are simple and well defined, the last one
remains a little obscure as to its actual structure within nervous system. We
analyze some experiments by Jim Gould to show that {\em what} is represented
as well as the {\em existence} of such representation by neural structures are
clearer. We distinguish liberty and freedom. Liberties can be found even in
physics but freedom concerns only the living organisms---even bacteria---, so
our language is a little different from the common use. The concept of freedom
includes structures and processes in living organisms by means of which they
take advantage of liberties. Several freedoms are described and the respective
structures are studied. We finish with the freedom of will, which turns out to
have the same general structure as all other freedoms. It is understood as a
natural phenomenon and given a position in the class of similar phenomena. We
specify the role of consciousness in this particular case and use it to give
the experiments by Libet a new interpretation.

Two processes underlying any freedom are the realization of the alternatives
contained in a liberty and the selection of one from the list of the
alternatives. Some kind of memory is always necessary to carry out the two
processes; hence use of memory is the third basic process. As a rule, the
realization contains a lot of randomness while the selection is mostly causal.
This is possible because the weak causality principle does not abolish the
causality wholesale.

To prove that the determinismus is wrong does not belong to the aims of the
present paper. We just accept that it has become very unplausible after eighty
years of quantum mechanics. Everything we do is to {\em assume} that the
determinismus is false and that the weak causality principle is true. Then, we
shall check the self-consistence of the new framework, see if the validity of
the principle and the existence of specific liberies are compatible with the
contemporary empirical knowledge as well as find the new theoretical ordering
and understanding of this knowledge provided by the new language.

\section{What the laws of physics do not bind}
We start with physics not just because the author is a theoretical physicist and has thus some advantage here but mainly to show that physics cannot be blamed for forcing determinism on us. Moreover, we shall observe that there is a lot of liberty in physics and shall be able to study the nature of the liberty in a rigorous way.

In the present section is also an attempt to explain the relevant physics in a way that can also be followed by non physicists. It will therefore be necessarily simplified and many details that are dear to the heart of a physicist will have to be skipped. We will also avoid all technicalities.

\subsection{Old but nor dead: Newton mechanics}
What is generally known as mechanics is called Newton mechanics here in order
to distinguish it from quantum mechanics. The main question of this section
is: how much is controlled and ruled by the laws of physics? If we are to
understand this more or less clearly, we need to understand the basic common
structure of the laws of the theories such as Newton mechanics, quantum
mechanics and general relativity. A few abstract and general notions are
needed for that: system, dynamical equation, state and space of states.

The system in Newton mechanics consists of particles with given masses. Only
after the system is chosen (the number of particles, their masses and
interaction with each other and with the outside), the theory give us its
dynamical equation, i.e, the law that every motion of the system must fulfil.
The law alone however does not determine the motion. To obtain a unique
motion, a state of the system must be chosen at a given instant of time,
mostly at the beginning of the motion and then evolved by the dynamical
equation. The state in Newton theory is constituted by the position, direction
of motion and velocity of each particle. All possible states form the
so-called space of states.

Let us consider an example. The trajectory of a bullet depends on the position
from which it is shot, on the direction of the gun and on the amount of powder
loaded. The position can be described by three coordinates (three numbers),
the direction by two angles (two numbers) and the amount of powder is
tantamount to the velocity of the bullet, which is one number. In this way,
six numbers are sufficient to describe the initial state of the bullet. The
resulting ballistic curve of the shot is unique in principle. It can be also
calculated from the dynamical equation. The bullet is then at some position,
it has one direction of motion and some velocity at every time instant after
the shot, that is exactly one state. As every state is given by six numbers,
the state space can be viewed as the set of all number six-tuples.

This is the general logical structure of the laws. What remains undetermined
by the theory? First, the choice of the system does. The choice determines
what is the dynamical equation and what is the state space. Then, any state
from the state space is freely eligible and the choice makes the motion
unique. Isaac Newton was aware of this feature. For instance, the fact that
all planets known to his time moved in the same plane around the Sun could not
be derived from his equations and he wrote \cite{Newton}: "Deus corpora
singula ita locavit." In our language, God has chosen the initial state.

The freedom in the choice of state is usually understood in a passive sense as
the generality of the dynamical equation, that is, its applicability to many
different situations that may occur spontaneously in Nature. It seems,
however, that one can go a step further and interpret the freedom as an active
freedom of physicists. The assumption can be formulated as follows.
\begin{quote}Physicists are free to choose a system from a broad system class
  and a state of the system from a large pert of the corresponding state
  space. Then, they can set out this system in this state in a laboratory (or
  elsewhere) at an arbitrary time.\end{quote}

We can call this hypothesis Realizability of Physical States. The liberty that
Newton attributed to God in the large is so attributed to physicists in the
small. There is a saying that everything not explicitly allowed is forbidden
in German-speaking countries while everything not explicitly forbidden is
allowed in English-speaking ones. Accordingly, we adopt the Anglo-Saxon
standpoint here.

I could not find a direct formulation of the realizability of physical states
in the literature. It seems that it is always tacitly assumed in the work of
experimental physicists. In any case it is completely compatible with
empirical praxis as well as with everyday laboratory work. Generally, this
kind of experimental freedom seems even to be one of the basic assumptions of
science. As concerns the trajectories of the bullets, the hunters have the
experience that they can carry their guns everywhere in order to shoot from
there in any direction they like. The hunt would not be much fun else.

One of the basic principles of statistical mechanics can be viewed as a
statement about a different kind of liberty. To explain this principle, let us
limit ourselves to thin gas in equilibrium. Such a system contains an enormous
number of particles. It is practically impossible to determine the state of
such a system by some measurements, or to realize a chosen state of it in the
lab. The available information about the system includes only values of some
overall quantities such as total energy, particle number and volume. There are
many states that are compatible with such description. Now, the principle that
we are explaining says that all states that are compatible with fixed energy,
particle number and volume are equally probable. (The name of this principle
is Micro-Canonical Distribution). More precisely, if we set up very many
vessels that have the same volume and that contain the same number of gas
particles with the same total energy in each vessel, then every allowed state
appears with the same frequency. Independently of how the vessels with the gas
are manufactured, all of the allowed states are present with the same
probability. Thus, we can realize any of them although we do not know and, in
fact, cannot find, which. Still, the principle has many interesting
consequences and is very useful.

\subsection{Rise and fall of determinism}
An important feature of Newton mechanics is that the values of all observable
mechanical properties of a system are uniquely determined by the state. In our
example, the energy, momentum, angular momentum etc.\ of the bullet to a given
time can be calculated from its position, motion direction and velocity at the
same time. Assume that the entire world is mechanical, that is, it can be
reduced in its entirety to a system of massive particles and forces between
them so that all properties of all objects could be calculated from their
mechanical properties, then a surprising consequence follows. If the state of
the world were known at some time, then everything what can be known about the
world at any other time could be calculated in terms of the world dynamical
equation. Even if no such complete knowledge or calculation were possible, be
it for practical or principal reasons, but if the world were a mechanical
system, then it would still follow that everything what ever happens including
every detail is predetermined (or post-determined). The only liberty that
remained would be the freedom in the choice of an initial state from the
(huge) state space of the world. This view of the world is called determinism.

Determinism was popular in the nineteen century because of the great progress
then in the project of reducing all physical properties to the purely
mechanical ones. For example, temperature can be so explained and calculated,
if one assumes that macroscopic bodies consist of invisibly small
particles---atoms or molecules---and that these particles move according to
mechanical laws. It turns out that the temperature of a body is proportional
to the average one-particle energy of its constituent particles. However, in
the first quarter of the next century, the more basic quantum theory emerged,
and this theory does not support the deterministic view (we shall study
quantum theory in the next subsection).

Newton mechanics, if cut down to size, remains valid. If we restrict ourself
to systems of macroscopic bodies that are not sensitive to the influence of
the quantum micro-world, that move with velocities that are much smaller than
the velocity of light, and if the gravitational field is weak then the bodies
would move with high precision according to Newton mechanics.

Now, an important aspects of the principle of realizability of physical states
from the previous subsection can be explained. There, a real freedom rather
than an apparent one has beed postulated. This would contradict the validity
of Newton mechanics if the principle would concern only mechanical systems.
The inclusion of physicists into the formulation of the principle allows us to
avoid the paradox, because Newton mechanics cannot be considered as valid for
the whole extended system consisting of the original one plus the physicist.

Although the freedom of physicics thus concerns the interpretation of some
aspect of Newton mechanics, it is primarily a property of living organisms.
This, it turn, cannot be studied by Newton mechanics and we don't understand
it yet. It will be studied systematically in Sec. 4.

\subsection{What we learn from quantum mechanics}
Quantum mechanics has the same basic logical structure as Newton mechanics.
Again, there is an affluence of various quantum systems. With each system a
dynamical equation and a space of states is associated. Given a state at a
time instant, then the state at any other time can be calculated from the
dynamical equation and is unique.

And again, the choice of system and state is not restricted by any rule in
quantum mechanics, only by practical feasibility. The only difference is that
the realizability of physical states is explicitly formulated in some
textbooks of quantum mechanics. For example \cite{Peres}, P. 48, contains the
realizability as a part of the so-called superposition principle. This may be
partially stimulated by the fact that a formulation of quantum mechanics
without observers is difficult. Still, the original reason for including
physicists into the realizability principle of Newton mechanics need not hold
here. If quantum mechanics is applicable to the extended system containing
physicists no contradiction has to result.

There are however other important features that make quantum mechanics very
different from Newton mechanics. We cannot explain all, but two of them will
play an important role later. These are the indistinguishability of quantum
systems and the statistical character of quantum mechanics. The first means
that quantum systems of the same kind such as all photons, all hydrogen atoms
or all molecules with the same compositions are utterly and absolutely equal.
Two products of some mass production factory may look equal, but they can be
recognized from each other, we can, e.g., make a mark on one, there is no
question which of them is here and which is there, etc. This is impossible
with quantum systems even to such an extent that any physically sensible state
of a system containing two particles of the same kind must be invariant with
respect to their exchange. Moreover, there is a relatively small number of
different kinds while systems of each kind occur in a huge number. This
reducibility of the micro-world to few absolutely equal building blocks has no
analogy in the classical world. It will turn out to be important to the
biology and to the freedom of living organisms.

The statistical character is not apparent at the level of dynamical equation,
which determines the states uniquely and is sometimes classified as
deterministic, but is rather associated with the state\footnote{This is in
  fact analogous to the statistical mechanics, where states can be identified
  with distribution functions. The dynamical law is the so-called Liouville
  equation and it also determines the distribution function at any instant of
  time uniquely if it is known at one. The values of measurable quantities
  however are not determined uniquely, only their probabilities can be
  calculated from the distribution function.}. Given a state, then there are
quantities the measurements of which always give the same value from some set
that is called spectrum of the quantities. One says that such a quantity have
a sharp value in the state. Most quantities however are said to have no values
in the state in spite of their measurability. That is to say, their
measurements give different results even if the state on which the
measurements are done remains the same. Only the distribution of these results
can be calculated from the state, i.e., each value from the spectrum has a
fixed probability determined by the state. Let us show typical details by
means of an example.

As our system, we choose a single photon. Of course, we can never observe a
single photon in the everyday life. What we know as light is always a cloud of
a large number of photons. To create a single photon requires a sophisticated
technique (which need not be described now). The photon can moreover be
created in a state, say, in which its momentum has a sharp value. According to
the well-known Heisenberg uncertainty relation, the position of the photon is
then totally "unsharp". What does this mean for the measurement of the
position?

The spectrum of position is the whole space. We can measure the position from
a subset of the spectrum by a photographic plate. The basic property is that a
single photon can create only a single black point on the plate if it hits it.
Hence, "totally unsharp" cannot mean a large smeared smudge on the plate made
by a single photon. Instead, it means that repeated measurements will result
in many black points and that the distribution of the points, after very many
measurements have been done, is uniform. The following interpretation can, may
be, help. The photon in our state does not have any position at all before
hitting the plate. The position is "created" only by its interaction with the
plate. The interaction cannot be controlled and gives different results in
each run. Of course, one cannot readily imagine some object without a
position; one of the main principles of some philosophical theories of
existence is that existing objects must have positions. However, there is no
difficulty for this philosophical principle: one can simply imagine that it is
the state of the photon that is smeared throughout the space. An ocean is an
object, the position of which is very extended indeed. Only, the physicists
prefer to speak of the photon as not having any position to saying that its
position is the whole space because the position is a specific well-defined
quantity in quantum mechanics and the possible values of this quantity are
points (in our language, its spectrum is a set of points).

Generally, quantum experiments look as follows. First, the experiment itself
consists of a number of runs. In each run, we obtain a single quantum system
(here the photon) from a source, which is some macroscopic apparatus. The
source is constructed in such a way that the photon obtained from it in each
run is in the same quantum state. The (macroscopic) arrangement of the source
and the measuring apparatus (here the plate), which is again a macroscopic
system, is the same for each run. The runs are performed at different times
and have therefore some time order. They can be performed at different places.
In each run, we obtain a certain value from the spectrum which can be read off
at the measuring apparatus (here, the black points at the plate). If the
experiment has sufficiently many runs then the distribution of the values
obtained is well approximated by the probabilities that are calculated from
the state according to rules of quantum mechanics.

In effect, everything done by an experimentalist is to manipulate and observe
some macroscopic devices. The account of the experiment can be completely
reduced to description of the behavior of macroscopic objects without omitting
anything indispensable for its understanding. It is the macroscopic structure
of the source that says the physicist whether it sends out photons or whether
it will rather be electrons, as well as what is the state of the particles.
And it is a macroscopic change of the measuring apparatus that disclose to him
which value of the measured quantity has been found. Hence, the
unpredictability is not just hidden in the micro-world without any relation to
our macro-world. It is the macroscopic behavior that is not always
predictable. More precisely, in the photon experiment, the quantum mechanics
does identify a cause for the distribution of the macroscopic black points at
the plate. That is because all photons that has been sent have had the state
of a sharp momentum. But it does not specify any cause of a particular run
giving this particular black point and not another one.

The reader can be embarrassed at the strange dichotomy of micro- and
macroscopic that we have assumed in this subsection. This is an old problem of
quantum mechanics that is not yet solved completely, but there are some
promissing ideas \cite{GH}.

\subsection{Causality principle}
Causality is an ancient assumption. For instance, Platon's formulation in
Timaeos is:\begin{quote}Everything that happens must happen because of a
  cause; for it is impossible that anything comes into being without
  cause.\end{quote} However, we have seen that the most basic of physical
theory today, the quantum mechanics, keeps silence about causes of something
that happens. How can this be explained?

Roughly, there are two possible explanations. The first is to keep the
causality principle and to assume that quantum mechanics is incomplete: the
causes do exist but are not captured by quantum mechanics. Indeed, the
causality principle cannot be falsified: if we do not see the cause of
something that happens, then we can always assume that the cause exists but we
do not see it. The second is to accept that the quantum mechanics is complete
and to abandon the causality principle: the causes do not exist. This is what
most physicists but not all underwrite. For example, Albert Einstein was
unable to accept the completeness.

If one accepted the incompleteness then one ought to propose a specific
alternative theory, in which the description of states is quantitatively more
detailed than in the quantum mechanics. In this way, the causes could be
described as differences in the values of some additional parameters, the
so-called hidden variables. Such a theory had to be necessarily more involved
than quantum mechanics but it must simultaneously reproduce all its measurable
results. In spite of great effort of many years, no such theory has been
constructed and no empirical support for it has been found.\footnote{There are
  deterministic models of some restricted kind (non-relativistic) of quantum
  mechanics such as the pilot-wave theory. They are, however, not suitable to
  be extended.} Moreover, it has been shown that the hiden variables had to
have very strange properties (action at a distance etc.).

Hence, accepting the completeness of the quantum mechanics is more plausible.
It does not imply that we have to abandon the causality principle altogether.
There certainly are causes for a vast number of events that happen. Everything
we need is to modify the principle:\begin{quote}Something that happens must
  happen because of a cause. The rest of what happens can, however, come into
  being without a cause so that there is a free choice between possibilities
  from a specific list. What has a cause, what is free and what are the
  possibilities lists is regulated.\end{quote} Let us call this Weak Causality
Principle. A model of such a regularity is quantum mechanics. The causes and
liberties are strictly regulated so that we always know what is predictable,
what is not and what are then the alternative possibilities. More everyday
model are the rules of chess. There are some rules according to which the
stones must be moved but there is, in every position, more or less freedom in
the choice of the move compatible with these rules.

Can the weak causality be included into a coherent picture of the whole world?
In particular, is it compatible with the rest of physics? We have seen that
what happens within the Newton mechanics satisfies the (strong) causality
principle, but we have also mentioned that the validity of the Newton
mechanics is limited and this removes possible contradictions. Quantum
mechanics, of course, is all right.

\subsection{A subtle change in the interpretation \\ of general relativity}
However, there is another modern theory called General Relativity, which
describes the world on the large scale. An important theoretical concept of
general relativity is that of space-time. This is a four dimensional space so
that there are four independent directions at each point, three space-like and
one time-like and the space-time includes all space points at all times. The
manifold carries the so-called space-time geometry that determines distances
and time intervals and it also carries matter. The general structure is
similar to that of the Newton theory in that it admits a number of different
space-times (=systems), that there are states filling up certain state spaces
that are different for different space-times and that there is a dynamical
equation.

The global character of general relativity is quite essential if we are to
compare it with Newton or quantum mechanics. There does not seem to be much
choice of the system now especially when the theory describes the whole world
for all times. We can still maintain that the space-rimes represent different
models of the universe. It is, however, difficult to require from physicists
to set up an arbitrary state of a whole world in their laboratory at an
arbitrary instant of time.

The character of states is different in general relativity in still another
way. They are associated with spacelike hypersurfaces in the spacetimes. A
hypersurface is a mathematical constructions rather than anything real. A
space-time can be represented by many different trajectories in the space of
states corresponding to different foliations by spacelike hypersurfaces of the
space-time.

There is no mark on the space-time that would distinguish the present instant
from all the other ones and there is consequently no difference between the
structures of past and future. The usual interpretation is to say that this
difference is purely subjective and that the present instant can be anywhere
depending on where is the observer, while the space-time is considered as an
observer-independent description of the total reality. The reality is thus
fixed for all times. This deterministic conception of world is called Block
Universe by some philosophers \cite{Mellor}.

However, the picture of time that follows from the weak causality is dominated
by an asymmetry between past and future. The future does not yet exist and
more possibilities are still open for it. The past is fixed, in principle,
because the choices are done at the present instant. Such an asymmetry is not
new in philosophy. Our ideas are similar to Popper's \cite{Popper1}. Thus, the
task to make the weak causality compatible with general relativity requires to
solve two problems. First, to see if the randomness can be included into the
dynamical equation and second, to specify, where do the choices take place.

To prepare the inclusion of the randomness, some words must be said about the
nature of the dynamical equation. Each space-time has two aspects: the
space-time geometry and the matter. An essential feature of general relativity
is that gravity and space-time geometry are two aspects of one and the same
structure. Then, because matter creates and influences the gravity, the
space-time geometry must depend on the matter. The equation that couples the
matter and the space-time geometry is called the Einstein equation. If the
model contains no matter, then the Einstein equation can simultaneously serve
as the dynamical equation of gravity.

Now, in constructing a universe model, we are free to choose various kinds of
matter and this can also be done in such a way that the evolution of matter
depends on some random variables taking values from a set of
possibilities\footnote{More precisely, the state equation of the matter
  contains some parameters such that any time dependence of the parameters is
  compatible with the Einstein equation, (satisfying certain conservation
  laws) and can be considered as a part of the evolution. Such a construction
  can easily be performed for homogeneous and isotropic cosmological models.}.
Then, the corresponding dynamical equation will not be deterministic. A state
of the universe at one instant of time together with its dynamical equation do
not determine its state at the next one unless the next values of the random
parameters are chosen. Such a simplified model shows that randomness can in
principle be included in general relativity.\footnote{It should be mentioned
  that some models containing classical gravity and quantum matter have been
  studied thoroughly \cite{Wald} but then the gravity source has been assumed
  to be the average value of its quantum behaviour. This is of course only an
  approximation valid in the cases when the average value describes all
  quantum possibilities in a sufficiently precise way. (Technically, this
  means that the mean quadratic deviation is negligible.)}

However, an instant of time is a space-like hypersurface that can be chosen
arbitrarily in the spacetime. If a relativist calculates an evolution, he
first chooses an initial hypersurface and, second, he specifies some rule,
called gauge choice, of how time is then to proceed along with the evolution
of geometry and matter governed by the dynamical equations. It can be proved
that the result is independent of the gauge choice. We shall try to require a
similar independence in the evolution containing random elements. For that, of
course, would be necessary that the choices are the same even if they are done
along different spacelike hypersurfaces.

One obvious way to do that is to assume that the choices are made locally.
Then, if two hypersurfaces intersect at a point, one can recognize if the
choices done along one of them at the point coincide with that done along the
other at the same point. Let us call these points, or some smeared version of
it, Local Presences. We assume that local presences have an objective, that is
observer independent, real existence, that is, things really happen at their
time and places. The local presences they are the only source of our evidence
about reality.

Some readers could ask, if we are to worry about realism. Is it not already
generally accepted that quantum mechanical evidence, especially after Aspect
experiments, definitely disproved realism (we use the notion of realism as
explained, e.g., by \cite{Russell})? The answer is that it is not generally
accepted and that a lot of research on this problem is being done. We shall
try to keep to the old-fashioned realism as long as it will be possible.

From inside our local presence, we can observe what is just happening further
away, for example at the Andromeda nebula. What we can see is going on within
some local presences there, which have to be arranged along our past light
cone and are shifted by some millions of years from our local presence because
of the distance and the velocity of light. We still assume that what we
observe there are some aspects of an observer independent reality as it was in
its time there.

However, what is the past? Clearly, the past exists only as a memory (i.e., a
specific arrangement of synaptic strengths in some brain) or other kind of
record that an observer, or a family of observers, can make about the
observations done within each of their progressing extended presences. Only in
this indirect way does the past have to do with reality. (Childhood memories
are subjective, the child itself in its time was real.)

The records are analyzed, compared and ordered: processed. This is an
important part of the game. Certain entities can be found that seem to be
always there (such as space-time events, specific classes of objects and
fields). For certain aspects of the entities, temporal and spatial relations
seem to be valid, for instance, the arrangement of space-time events into a
smooth manifold with some geometry. Some causal relations can be summarized
and generalized in the form of evolution laws. Other aspects of the entities
can exhibit a kind of liberty. In this way, a picture of some broader
space-time structure emerges so that the pasts of different observers can be
included into a unique one. The aim is to construct a logically coherent set
of explaining and ordering hypotheses from which all evidence can be logically
deduced.

The past as a (processed) record seems to be fixed in all aspects and details.
There are two very different reasons for that. First, the choice from the
alternatives of all liberties has been done and no change is any more
possible. Second, we usually suppose that different observations or
observations of different observers concerning these already done choices can
finally be put into agreement, or that their contradictions can be
satisfactorily explained. In particular, any small neighborhood inside of a
past describes what happened when it was a presence and it can be considered
as a local presence of any observer being then in it, and the assumption is
that his observations within this local presence will not in principle
contradict ours. This is a rather non trivial hypothesis on which, in fact,
all of the science is based; it has a natural explanation in the philosophical
realism. We call this hypothesis The Uniqueness of History.

Finally, what is the future? The very existence of future is a hypothesis
based on the analysis of the records which confirm that, as yet, the presences
have always progressed. Similarly, we can extrapolate the existence of the
entities and the validity of the laws to where we cannot make direct
observations, in particular into the future of each respective cone. Only in
this way, we can make predictions. On the other hand, the predictions can
concern always only a part of the future. As we have seen, there are also
unpredictable aspects. Thus, some part of the world is newly created ("chosen
by Nature" under more possibilities, cf.\ \cite{Popper1}) at the presences,
another part is determined by the past.

The ideas described up to now are also supported by the contemporary knowledge
of how brain neocortex works, even in mice. It constructs a structured, i.e.,
already processed, memory record of all experienced (interesting aspects of)
presences and uses this material to create expectations (cf.  \cite{Hawkins}).
Similarly, the human science is being made, at least in principle and in rough
features, analogously, leading from records to predictions, too. The nature of
scientific reasoning is described in \cite{Jaynes}. We can say pointedly: What
really exists are only the local presences. The past as well as the future are
nothing but products of neocortex.

Now, we can answer the question of how the space-times of general relativity
are to be interpreted: The space-time must be just a hypothetical past, that
is, the unique history of an evolution assumed to be completed (cf.\
\cite{Ellis}). With other words, a spacetime can be viewed as one possibility
of the Liberty of Universe, which is the union of all liberties. The problem
of asymmetry between future and past does not even arise because we are
considering only the pasts.

The above is only a subtle reinterpretation of general relativity because it
does not seem to lead to changes in any calculation and any discussion
concerning measurable properties within general relativity and with the
Einstein equation done as yet. The reason is, that such calculations and
discussions can primarily apply only to past evidence and hypotheses formed
primarily about the past, as it has been explained above. If we accept this
change in interpretation of general relativity, then the weak causality
principle becomes compatible with the whole of the contemporary physics.

\section{How liberty can be defined}
The discussion of the foregoing section has already suggested the conception
of liberty as a choice among different possibilities that is compatible with
the laws of physics. To see the existence of such a liberty is still not easy
because of what we have called the uniqueness of history. The records of the
past are unique and hence there does not seem to be any freedom. Even if there
is such a freedom, the possibilities have already been chosen and the history
cannot be changed. How did we come to think that there is any freedom?

Recall how the described experiment revealed the liberty in the position of
the photon. The experiment consisted of many runs performed at different
times, each of them giving a different result. The conditions of each run were
specified so that the experimentalist could say: Each run started under the
same relevant conditions. This is the crucial point. Apparently, the time and
location of the run does not belong to the relevant conditions. Then, it
becomes meaningful to say that the same experiment is repeated and that it
gives the same or different result as a previous one. This motivates the
following definition:\begin{quote} The liberty of a system is associated with
  certain reproducible conditions and it is defined as the list of different
  possibilities that are open to the system under the conditions.\end{quote}
The important words "system", "reproducible conditions" and "possibilities"
have here a more general meaning than in the photon experiment and are
explained below.

The two words "liberty" and "freedom" will distinguish two different concepts
in what follows. The liberty as defined above is a relatively simple notion
that can be applied even to photons. The freedom will be applicable to living
organisms and will denote the fact that the organisms are equipped with the
structures, mechanisms and methods that enable them to utilize liberties
\footnote{Thus, the meaning of both words is appreciably extended in
  comparison with their current use. I apologize for this violence, but I
  could not find better words and shall accept any better proposal.}.

The term "system" need not be a simple physical system such as a photon or a
bullet, but can denote more complex objects such as living organisms. The
specification of the object that appears as the system in the definition can
be a part of the relevant conditions. This has been the case in the photon
experiment, where the nature of the source has constituted a part of the
conditions and guaranteed also that what has been sent out was a photon.

The term "reproducible conditions" expresses the main idea of our definition.
A liberty is always understood in connection to certain conditions. In
principle, broader or narrower conditions allow more or less liberty. However,
the choice of conditions is not arbitrary. We assume that the same set of
conditions is often fulfilled in different cases, may this happen
spontaneously in Nature or may it be sufficiently easy to be arranged by
people. A complete list and a clear description of the relevant conditions
must enable the check whether the same conditions are satisfied in different
cases or not. The reproducibility is the property that makes the liberties
empirically manageable and theoretically derivable, similarly as the laws of
Nature are.

The "possibility" ought to really exist as opposed to a purely thought one, in
the sense that its realization can be observed at least in some cases in which
the conditions are satisfied. It can be an effect of a cause that lies outside
the condition set so that some of the possibilities has its own cause in a
given case. It can as well be that for its particular realization no cause can
be found and even need not exist, such as it has been assumed in quantum
mechanics and stated by the weak causality principle. Or there can be a
mixture of chances with causes. Our definition of liberty is such that it does
not include the way in which its possibilities are realized. It may go in a
random way by a chance or in a deterministic way by a cause. The same
possibility of a given liberty can be realized in different ways in different
cases. We shall see complex examples with a lot of interesting structure
later.

The number or some other measure of the amount of all possibilities can even
serve as a numerical value of liberty. For example, if there are $N$ different
possibilities, we can define $\ln N$ to be the value of the
liberty.\footnote{If each possibility has its own probability, then the Shanon
  formula for entropy could be used.}

Let us recall some examples from the physics section. The conditions of the
liberty observed in the photon experiment are that the nature of the source is
to sends out photons in the state of sharp momentum, second, that the
measurement apparatus is a photographic plate of certain kind and third, that
the devices are arranged in a fixed way. All conditions concern only
macroscopic properties of the devices, but the standard interpretation of
quantum mechanics considers them as maximally narrow: no further conditions
exist that would be relevant. That is; other possible accompanying
circumstances such as e.g., the time and location of the measurement, the
conjunction of planets and stars, the mood of the boss, the state of the stock
exchange, etc. can indeed be shown to have no observable influence on the
course and results of the experiment. The possibilities are the points at the
plate that can become black. These are all pooints of the plate, forming in
this way a well defined list of possibilities. The liberty could be measured,
e.g., by the logarithm of the area of the plate.

Some liberties are important and useful even if their conditions do not form a
maximally narrow sets. That is, further conditions could be added, at least in
principle, so that the number of possibilities will decrease. Such liberties
do not logically contradict the (strong) principle of causality or the
determinism. The statistical physics of thin gases in equilibrium yields an
example. The conditions are that the total energy, the total volume and the
total number of molecules in the vessel have certain values at certain time
$t$. Such conditions are compatible with a huge number of mechanical states of
the gas molecules at $t$. The number of possibilities equals the number of
possible states. If any such state really occur at $t$ then there can be a
cause of it in the past to $t$ that has nothing to do with the conditions.
These conditions could in principle be narrowed so that just one arbitrary
fixed mechanical state would be allowed at the time $t$. Then, there would be
only one possibility.

Another liberty of such a kind is connected with the so-called emergent
phenomena. These are properties of complex systems that cannot be derived
exclusively from the properties of its individual constituents. One can also
say that the whole is greater than the sum of its parts. The simplest example
are two electrons, two protons and two neutrons. They can form either an atom
of helium or two deuterons so that we cannot say what are the properties of
the composition if no additional information about the structure is available.
The reason is that there are two possibilities of how the constituents may
combine. The possibilities constitute a liberty that we can call {\em liberty
  of combination}. We can generalize it by counting also the numbers to the
possibilities. Then, the three kinds of constituent above can combine into
about one hundred stable atoms and these atoms can combine into zillions of
stable molecules, crystals and mixtures. This is a huge liberty, which
underlies the surprising wealth of structures in Nature.

\section{How living organisms \\ take advantage of liberties}
The limited knowledge of the author in the fields of biology, ethology and
brain research may make the following deliberations somewhat uncertain. He
apologizes for irritating inaccuracies that a specialist surely would find if
any happened to read this section. Still, it seems that the main idea ought
not to be completely wrong in particular also because it is little more than a
reformulation of Darwinism built on the notion of liberty. We shall see how
this notion can throw some fresh light on a number of facts of life.

\subsection{Liberty of mutation}
Let us start by a short story about how a species of bacteria called
staphyloccocus aureus develops a resistance to a new antibiotic. The
antibiotic reacts with a molecule of the bacterium cell in such a way that
some life process is disturbed. Thus, this molecule ought to be changed so
that it does not react lethally with the antibiotic any more. Of course, the
structure of the cell must be sufficiently flexible so that all life processes
can also run with the new molecule. Let us call mutations all changes that
satisfy the second condition. The allowed mutations can be considered as the
possibilities of a liberty. We call it the liberty of mutation. All conditions
of this liberty is just the ability to live of the mutated staphiloccocus
cell. The resistance can develop only if the liberty of mutation is
sufficiently large.

How do mutations come about in the first place? It seems that this is more or
less random process that works all time and that has no plan or aim.
Bombardment by some radiation such as cosmic rays, disturbance by some
contingent chemistry and physics, or even some contact with other bacteria and
viruses can be effective. Mutations occur with single molecules and it seems
that quantum mechanics is important for them. In any case, there is enough
space for randomness. It is true that the occurence of the mutation protecting
the bacterium from the antibiotic has non-zero but very low probability.
However, there is a very large number of staphyloccocus around. This number
multiplies the probability, making the favorable mutation feasible.

The cell with the mutation must be there before the antibiotic is applied. The
antibiotic then kills all cells except for those that exhibit the advantageous
mutation. We can say that the antibiotic makes a selection, a choice among the
possibilities of the liberty.

Finally, the trick must be remembered in some way so that it can be applied
against each future antibiotic attack. The mechanism working in bacteria is
the following. Every cell has a genetic blueprint, written down in a
particular molecule of deoxyribonucleinacid (DNA). The mutation must first
appear in the molecule of DNA of a parent cell, and only then, as the result
of the cell division, it is referred to the molecule that reacts with the
antibiotic in a daughter cell. That is because the new blueprint is used for
the construction of new cells in the process of cell division. Each of them
also inherits a copy of the blueprint. In this way, the mutation is completed,
remembered and proliferated.

DNA is a chain of four kinds of building blocks, the nucleotides. The ability
of DNA to carry information is based on the indistinguishability of different
nucleotides of the same kind and so on quantum mechanics. The most important
property is that the sequence of nucleotides can be arbitrary. Each sequence,
of any length and order, can be joint into a stable molecule of DNA. This is a
kind of chemical liberty with the possibilities being the different sequences,
and it is restricted only by the problem of keeping very long chains
undisturbed and accessible.

We can identify three essential processes in how living organisms utilize
liberties. First, there must be a real liberty in our sense, that is, its
conditions are fulfilled sufficiently often and all its possibilities can
really occur or, they can be realized; we call this process realization. The
possibilities are realized in processes including pure chance, contingency as
well as causal laws. The liberty must be sufficiently large to contain some
advantageous possibilities. Second, the choice between the possibilities must
be done so that the advantageous one prevails; we call this process selection.
Third, the advantageous possibility must be remembered for future use; we call
this process memory. Realization, selection and memory are very general
processes that always constitute the strategy of living organisms with respect
to liberties. More exactly:\begin{quote} A Freedom is the ability of living
  organisms to carry out the processes of realization, selection and use of
  memory with respect to a liberty.\end{quote}

An example of how the realizability concerning the mutation liberty can be
improved in complexer organisms than bacteria is the phenomenon of sex. The
claim that sex would enhance human liberty may seem preposterous. However, we
have in mind the liberty of mutation rather than the freedom of will.

Let us give a simplified introduction to sex phenomenon that will be
sufficient for understanding which liberty is improved by sex and how it
works. We will draw upon \cite{Dawkins} to a large extent. First, within the
whole genetic material of an organism, its DNA, shorter pieces called genes
can be found each of which code for some property, such as the color of eyes,
say. Within all individuals of a fixed species, more genes with the same
function but different results can be found. In this way, more genes, like the
brown eye and the blue eye gene, are rivals for the same slot on the DNA; such
rivals are called alleles. The origin of the alleles lies in a step by step
mutations occurring in different lineages. There is a sense in which the genes
of the population including all alleles resulting in this way can be regarded
as a gene pool. The population is constituted by all contemporary individuals
of a species. In this sense, the pool is a propriety of a species at a given
time.

Now, there are many ways in which a possible gene combinations forming a whole
DNA molecule that would encode for a viable individual could in principle be
chosen from the pool. All these combination possibilities form a part of what
we have called the liberty of mutation. But could such combinations come about
in a reasonable time? It turns out that the phenomenon of sex does just that.

Each cells of an individual contains the blueprint for the whole body, not
only for the cell itself. In the species that can reproduce sexually, most
cells of an individual contain exactly two copies of it, one from the father
and one from the mother of the individual. Only the sexual cells of the
individual, eggs or sperms, contain just one copy. During the manufacture of
these cells, some bits of each parental DNA physically detach themselves and
change places with exactly corresponding bits of maternal DNA. The process of
swapping bits of DNA is called crossing-over. It seems that the choice of the
points on the DNA where the pieces have their ends is random. It is, moreover,
different in each sexual cell of the same individual. The density of the
points at which the DNA is broken by crossing-over is sufficiently large for
something to happen at all and sufficiently small so that there is a large
probability for clusters of several genes to stay together and to be copied
truly.

Because of sex and crossing-over the gene pool is kept well stirred, and the
genes partially shuffled. Thus, the realization of very different
possibilities of mutation liberty is accelerated so that the incidence of bold
changes is strongly enhanced. In the whole process starting from the choice of
sexual partner through the crossing-over in each sexual cell to the
combination of a sperm with an egg, something is subject to causal laws but a
lot is purely accidental. This is often so with the realizability.

The selection mechanism works only on the level of individuals because it is
driven by the success or failure of whole individuals. This means that the
selection does not act on the genes directly. As far as a gene is concerned,
its alleles are its deadly rivals, but other genes are just a part of its
environment. The effect of the gene depends on its environment. Sometimes a
gene has one effect in the presence of a particular other gene, and a
completely different effect in the presence of another set of companion genes.

The memory that would be necessary to remember the best combinations of
alleles is worsened by the sex. The necessary random break up of the whole
combination comes about independently of how advantageous the DNA of the
mother or of the father has been. Only those pieces of DNA can be copied truly
that are short enough so that their break up during crossing-overs has a very
low probability. These can be clusters of just a relatively small number of
genes. That is why single genes or relatively small clusters of genes are
units of heredity in the sexually reproducing organisms rather than the whole
DNA as in bacteria.

Hence, the long-term consequence of non-random individual death and
reproductive success are manifested in the form of changing gene frequencies
in the gene pool. Evolution is the process by which some genes become more
numerous and the others less numerous. On one hand, sex greatly improves the
realizations, on the other, it subtly impairs the memory. Sex is a delicate
phenomenon.

\subsection{The liberty of motion}
Some multicellular organisms such as animals possess an additional liberty
that we shall call liberty of motion. This means that parts of animal body,
e.g., trunks, legs or wings, can take different relative positions to each
other without inhibiting other functions of the body. The change of this
relative position, if there are no external hindrances, can be carried out
with various velocities and external bodies can be shifted thereby with
various forces. This defines a list of possibilities---the liberty---that can
in principle be realized by each individual body. Plants can also perform
limited motions and have some choice, for instance between different
possibilities of growing, but these are not included in our definition.

It seems that animal motions are always organized with the help of some
nervous system. Experiments show that certain nerve signals trigger certain
motions. "Useful" sequences of motions such as running or flying are carried
out by specialized sets of nerves connected in a particular way. Moreover,
some influence of sense data on motions are made possible by other connections
of nerves. The influence can be direct so that some stimulus elicites some
motion, or modulary so that some stimulus modulates the strength of some
direct connection.

The wiring of the nervous system---which neuron is connected to which---is
inherited and fixed. However, the strength of some part of the connections is
variable. There is a liberty called synaptic plasticity \cite{Kandel}. Its
possibilities are different strength of the connections. There is a certain
threshold so that the connection is broken, if the strength falls under it
etc. The animal can start with arbitrary strengths; the starting strengths
contain an some space for randomness. The choice from the list of
possibilities is causal, certain choices following certain stimuli or
(time-ordered) sets of stimuli. Such changes in the synaptic strengths can
last several minutes, which constitutes a short-term memory, or for weeks or
years, the long-term memory. The process is called learning. Many examples of
learning starting with invertebrates are can be found in \cite{Kandel}.

The learning has some limitations, because many connections are fixed. An
example thereof is the experiment with the insect species sphex ichneumoneus
as described by Dennett \cite{Dennett}, P. 82: \begin{quote} When the time
  comes for egg laying, the wasp Sphex builds a burrow for the purpose and
  seeks out a cricket which she stings in such a way as to paralyze but not
  kill it. She drags the cricket into the burrow, lays her eggs alongside,
  closes the burrow, then flies away, never to return. In due course, the eggs
  hatch and the wasp grubs feed of the paralyzed cricket, which has not
  decayed, having been kept in the wasp equivalent of deep freeze. To the
  human mind, such an elaborately organized and seemingly purposeful routine
  conways a convincing flavor of logic and thoughtfulness---until more details
  are examined. For example, the wasp's routine is to bring the paralyzed
  cricket to the burrow, leave it on the threshold, go inside to see that all
  is well, emerge, and then drag the cricket in. If the cricket is moved a few
  inches away while the wasp is inside making her preliminary inspection, the
  wasp, on emerging from the burrow, will bring the cricket back to the
  threshold, but not inside, and will then repeat the preparatory procedure of
  entering the burrow to see that everything is all right. If again the
  cricket is removed a few inches while the wasp is inside, once again she
  will move the cricket up to the threshold and re-enter the burrow for a
  final check. The wasp never thinks of pulling the cricket straight in. On
  one occasion this procedure was repeated forty times, always with the same
  result.\end{quote} What appears here as a lack of freedom is in fact a lack
of the freedom precisely in our sense: some connections cannot be changed.

\subsection{Liberty of portable neural representatives}
There are processes in nervous systems that are based on learning but are more
complicated than it. The existence of such processes in nervous system of
invertebrates are suggested e.g. by experiments by Jim Gould with honeybees.
The following description is borrowed from \cite{Hauser}.

The bees seem to remember some aspects of the environment of their hive and
are also able to describe routes within this environment to each other by a
kind of body language, the so-called dance. \begin{quote} ...finding food
  depends less on luck and more on sampling from relatively well known
  foraging sites, areas where food availability depends on seasonal variation
  in polen. When a honeybee forager returns and dances, other hive mates pay
  attention. Depending on information in the dance and the current needs of
  the hive with respect to finding food as opposed to storing it, the
  observers will either stay put or go out on their own foraging expedition.
  The observer must therefore process the information in the dance and then
  place it within a system of spatial representation...

...Gould observed a hive that has been maintained near a lake for a long
period of time. This provided some insurance that the honeybees were familiar
with the local environment. Each day, one group of foragers was trained to
move from a release spot away from the hive to a boat on land, stashed with
nectar; once they loaded up on a meal, they were captured and prevented from
returning to the hive. Over the course of several days, the boat was displaced
further and further from the release site until one day it was square in the
middle of the lake. At this point, the foragers were allowed to collect nectar
from the boat and then return home. When the foragers arrived at the hive,
they danced, indicating the location of the nectar-ladden boat. Although the
hive paid attention to the dance, virtually no one flew out of the hive. Gould
suggests that the honeybees responded to the forager's dance by referencing
their cognitive map. As for this colony, the map fails to reveal a "Food Here"
sign in the middle of the lake. Sceptical of the dancer's message, hive
members wait for a more reliable dancer. (PP. 77-78.)\end{quote}

The information about the environment is represented and stored in the
honeybee nervous system forming thus a real entity different from the
environment itself. It has to be created during the life of individual
honeybees rather than built in from the inherited DNA (to build it into the
DNA would require thousands of years): it has been learned. In Gould's
experiment, some neural representative of the position of, or the way to, the
food source forms in the forager honeybee nervous system from the sensory data
and its own motions during the flight; it is also learned. The dance
reexpresses it as a sequence of motions that can be "understood" by the hive
mates. That is, observing the dance they can build, learned in this way, a
representative of the food way in their nervous systems.

The representatives of sensory data, of motion sequences and of environments
concerning Gould's honeybees are examples of what we shall call portable
neural representative, (PNR). It is a neural representative formed during the
life of an individual and the nervous system is able to work directly with
these representatives; the word "portable" is to distinguish it from the fixed
neural representatives of, say, sequences of motions that have been inherited.
Any PNR must clearly be based on some arrangement of synaptic strengths in a
specific set of neurons, which are not known in detail. The existence of such
a representative is just inferred from behaviour. The name "portable neural
representative" reminds of what is really known now. It is, however,
sufficiently general, it leaves many details open, and is thus suitable for
our purposes.

The bee nervous system is then apparently able to compare the message PNR with
the PNR of the environment in order to see whether one is to stay put or to
fly out. These and other PNR (such as concerning some work in the hive) are
apparently used in the process of selecting the motion sequences before any
actual motions are done. The chosen motion is not just a learned specific
response to a specific cue. There never has been such a dance before!

Important observable data in this experiment are the numbers of bees that fly
out and that do other motions. The relative numbers differ from case to case.
The bees themselves are identical clones from the point of view of heredity.
This suggests that the different decissions must be due to different PNR to
start with. If so, there may also be some random factor.

More numerous and more convincing are experiments with mammals showing similar
or more flexible use of memory for choices of motion \cite{Kandel}. The
honeybees are much more primitive, in particular they do not posses any
hypocampus. There is some discussion about interpretation of Gould's
experiments \cite{Hauser}. However, for the thesis of the present subsection,
Gould's experiments are not vital.

What is the relevant liberty? The ability of nervous systems to form in
principle more or less arbitrary PNR without disturbing the function of the
system is similar to the liberties of mutation or motion and we call it
liberty of portable neural representative. It is apparently based on the
synaptic plasticity However, some more specific account of this liberty
similar to the previous two is difficult because little is known about the
actual structure of PNR in nervous systems. Neural network models of nervous
systems might probably be used to get some insight. In any case, this part of
Sec.\ 4 is more hypothetical than the previous one.

This experiment shows how the memory enters the process at many points, we can
also find a well-defined selection, but we have only a nebulous idea of how
the realization of the possibilities and how the selection proceed within the
individual nervous system.

After the choice of a PNR for going out or
staying put, the nervous system brings about an actual sequence of motions
compatible with it. Thus, the liberty of PNR is associated with the liberty of
motion in a similar way as the liberty of mutation is, but it constitutes a
distinctly different kind of liberty. By it, the choice procedure is shortened
from the time interval covering many generations to a time interval shorter
than one individual life. In such a way, the nervous system that might
originally just serve to organize motions into suitable sequences becomes the
most powerful instrument for utilizing liberties by living organisms.

\subsection{The freedom of will}
The freedom of will is usually understood as the freedom to select in mind
consciously an idea of an action and then to carry out the action. The action
can be a sequence of motions, but it also can be another conscious mental
action. The consciousness component distinguishes this freedom from the
others. We assume that every idea has a portable neural representative and
that it is conscious. Then we can effectively restrict ourselves to PNR and
the liberty underlying the freedom of will is again the liberty of PNR .

What is exactly the consciousness, in particular, how it is represented by any
nervous processes in the brain, seem to be not known. However, some
phenomenological understanding is possible and it will be sufficient for our
purposes. This means that whether or not anything is conscious for a person
must be decided or reported by the person \cite{Libet}. Of course, recent
findings support strongly the idea that unconscious processes are very
important and ubiquitous in all cerebral activities. Accordingly, the role of
the freedom of will must be smaller that some people would like to think. But
we are going to argue that the existence of a kind of this freedom can be
assumed without problems.

To begin with, it may be interesting to observe that consciousness has a
strong memory component. Not only is one aware of something (that is,
conscious of it), but one must also be aware of what one was aware of in the
past, so that the well known roughly continuous, time ordered, stream of
consciousness results. This so called declarative memory is heavily used by
consciousness all the time.\footnote{However, the process of getting directly
  aware of something is known to be independent from the process of storing it
  in the memory.}

The crucial hypothesis of this subsection says that the consciousness is a
tool that enables one to better use one's liberty of PNR because it is
essentially an instrument for complicated symbol manipulations or scenario
runnings as well as for the utilization of such calculation results for the
choice of actions. The unconscious brain is usually not able to do such
calculations alone because it has originally had a different purpose and this
may partially explain that all conscious thinking is rather awkward,
energetically expensive and relatively slow.\footnote{The consciousness may be
  slow for other reasons, too \cite{Libet}.}

It is interesting that digital computers are also tools to make complicated
calculations. Thus, the consciousness can be understood as a method by which
we make our brain emulate a digital computer. This aspect of consciousness is
well-known and a relatively simple one. It is central for the understanding of
the freedom of will that will be put forward in this paper. We can leave the
question open, whether or not the consciousness is, at least in main features,
reducible to this aspect.

This is not to say that brain solves complicated mathematical problems in the
same way as a computer would do. Actually, we do not maintain that unconscious
processes are excluded from the whole solution process. Just the opposite is
well known to be true. However, conscious processes are necessary at some
stages similarly as computers are needed at some stages of modern research
projects.

\subsubsection{The experiments by Libet}
Our hypothesis has several agreeable properties. One of them is that it gives
a nice and unexpected interpretation to the celebrated experiments by Benjamin
Libet \cite{Libet}. Libet's experiments are ingenious and very enlightening
for everybody studying consciousness, but we shall not accept all Libet's
interpretations. A short account of relevant material that will be sufficient
for our purposes can be found in the Foreword to \cite{Libet}:
\begin{quote}Libet's work has focused on temporal relations between neural
  events and experience. He is famous in part for discovering that we
  unconsciously decide to act well before we thing we've made the decision to
  act. ... Libet asked people to move their wrist at a time of their choosing.
  The participants were asked to look at moving dot that indicated the time,
  and note the precise time when they decided to flex their wrist. The
  participants reported having the intention about 200 milliseconds before
  they actually began to move. Libet also measured the "readiness potential"
  in the brain, which is revealed by activity recorded from the supplementary
  motor area of the brain (which is involved in controlling movements). This
  readiness potential occurred some 550 milliseconds before the action began.
  The brain events that produced the movement thus occurred about 350
  milliseconds before the participant was aware of having made the decision.
  Libet shows that this disparity is not simply due to extra time required to
  note and report the time.\end{quote}

Let us compare the experiment by Gould with that by Libet. In both cases, it
is the nervous system that selects some PNR and starts the action. In both,
the liberty of PNR and of motion is utilized. The two freedoms seem to be of
the same.

From the evolutionary (that is, natural selection) point of view, the
conscious component in selecting PNR carries with it both advantages and
disadvantages. On one hand, it enables complicated deliberations and
calculations, on the other, it consumes a lot of energy and time. Now, if we
look at the action asked for in Libet's experiment, it itself does not require
any calculations, it is a simple choice of time instant. Still, consciousness
has been used by the participants before the experiment in order to understand
the task, to reduce it to the simple choice of time instant and to prepare it
thus for the performance. Then, when the experiment is running, it seems
natural that the unconscious brain does not switch in the consciousness
because no complicated scenarios are to be elaborated. It is satisfied, after
it has done the work itself, with merely dropping a notice to the
consciousness to enable a possible veto, and to the consciousness journal
(declarative memory) to save this information for possible later use. It seems
that the consciousness is not the master but a servant.

Our conclusion from Libet's experiments is therefore different from that of
relatively many philosophers or natural scientist. They seem to find there a
strong suggestion, or even a proof, that there is no freedom whatsoever and
that one's impression of having some is just an illusion. We want to maintain
that there is a lot of freedom, even, say, for the honeybees and that human
freedom is even larger because people can find more possibilities for PNR with
the help of conscious calculations. In our language, the realizability is
enhanced. In effect, we do assume that freedom is not an illusion, but we give
the consciousness a smaller role in it.

\subsubsection{An example: playing chess}
Playing chess is an activity that clearly shows the value of conscious
calculations. There is a well defined liberty: all moves that are allowed by
the rules of the game in a given position. Moves are understood as ideas, not
as actual motions of the stones (a move is a whole class of such motions), and
we again assume that each move is some conscious PNR.

As a beginner, one is happy to do the first move which pops into one's
consciousness in each position and looks promising, but that leads mostly to a
disaster. What is to do is to consciously calculate developments to which such
an idea would lead without carrying out the moves on the chessboard; a number
of such developments quickly grows if the analysis is extended to the depth of
three or more moves and the situation becomes rather messy. Thus, it is no
miracle that many digital computers play chess better that most people.

The selection between possible moves is done according to the purpose that one
is following. Even if one just wants to win, one's choices might be different
in the same positions because of changing skills or because one adapts the
choice to one's knowledge of one's opponent. It is also conceivable that one
does not need to win. For example, one would like to teach one's child to play
the game, etc. In any case, there is a reason for one's choice and we can say
that each move that is actually performed by a given player has a cause within
the player, at least in regular circumstances.

This idea is rather similar to the philosophy of compatibilism, which
attempted to make freedom of will compatible with determinism (see, e.g.,
\cite{Dennett}). Although we have rejected determinism, this particular idea
of compatibilism is fully taken over here. Hence, the essence of our freedom
is not in the randomness of our moves. Of course, we can decide to make our
moves with the help of dices, but this possibility does not exhaust the
concept of our freedom. We agree with the compatibilists that the player is
the cause of the player's moves but we do not follow them further in excluding
that the corresponding causal chain has started within the player.

It is only when we are getting at a sufficiently large pool of trial moves
from which the actual move is to be selected where the randomness often plays
a role. The problem is that the computer of our consciousness is not able to
do a systematic analysis even in chess. (Today's digital computer cannot
calculate the game all the way to the end, either.) We are therefore looking
for some move that is motivated by some properties of the position over which
we are sitting. Different moves occur to us in a way that is not completely
systematic. Trying to calculate possible consequences of each, we learn more
about the position so that after rejecting one idea, we are likely to get
another, etc.; it is the method of trial and error (cf.\ \cite{Popper2}). work
need not obey any overall algorithm. Algorithmic calculations of our conscious
mind's are mixed with some input of the unconscious brain. Even if the
unconscious brain does not calculate, it often provides surprising
associations of ideas, which themselves may but need not have been obtained by
calculation. It seems that the way we arrive at the trial moves does not form
a strictly causal chain\footnote{There is some analogy here to the way the
  mutation liberty is used by bacteria: the trial ideas are similar to
  mutations that come about in a partially random way and the selection in
  both cases is causal.}. In spite of the unconscious component of getting
trial ideas, the definitive move selection is, as a rule, conscious and it is
indeed the cause of the move that is actually carried out.

The final point
is that after calculating through the scenarios and choosing among them
consciously, the act of actually moving the stone may contain unconscious
elements. The choice exactly when, how rapidly and along which trajectory the
move is to be done can be, and mostly is, done without any consciousness.
However, this is another liberty. The liberty of the choice between possible
chess moves is used with the help of the conscious calculator. The choice
between possible stone moves compatible with a given chess move is a different
one, which is not relevant to our problem. The question if the conscious mind
directly moves the body hand seems to be less important than whether the body
hand moves in accordance with the mind plan or not.

The game of chess is,
of course, a strongly simplified model of life. It provides, however, all the
relevant features of conscious decisions and gives thus an example that fits
the above definition of the free will.

To summarize, the notion of freedom that is defined here is not an illusion.
Still, it is a more complicated phenomenon than is usually assumed. The whole
concept of freedom of living organisms must be separated into several
different notions. First, a kind of relevant liberty must be identified, e.g.
the liberty of motion. This is a relatively simple concept that can be
described in precise terms and that can be studied experimentally. Second,
there are ways and methods of how organisms make use of the liberty. For this
an involved, many levels (DNA, nervous and locomotive systems) structure
evolved which supports both random (in realizations) and causal processes (in
selections) and includes a relatively large memory. The way it works is again
accessible by experimental study.

\end{document}